\documentclass[prl,twocolumn,showpacs,superscriptaddress,preprintnumbers]{revtex4-1}
\usepackage{amsfonts, amsmath, amssymb, bm, bbm, graphicx, epsfig, epstopdf}
\usepackage{dcolumn}
\usepackage{textcomp}
\usepackage{units}

\newcommand{\ket}[1]{|{#1}\rangle}

\begin{document}

\title{Ultrafast slow-light: Raman-induced delay of THz-bandwidth pulses}

\author{Philip J. Bustard}
\thanks{These authors contributed equally to this work.}
\address{National Research Council of Canada, 100 Sussex Drive, Ottawa, Ontario, K1A 0R6, Canada}
\author{Khabat Heshami}
\thanks{These authors contributed equally to this work.}
\address{National Research Council of Canada, 100 Sussex Drive, Ottawa, Ontario, K1A 0R6, Canada}
\author{Duncan G. England}
 \address{National Research Council of Canada, 100 Sussex Drive, Ottawa, Ontario, K1A 0R6, Canada}
\author{Michael Spanner}
 \address{National Research Council of Canada, 100 Sussex Drive, Ottawa, Ontario, K1A 0R6, Canada}
\author{Benjamin J. Sussman}
\email[email: ]{ben.sussman@nrc.ca}
 \address{National Research Council of Canada, 100 Sussex Drive, Ottawa, Ontario, K1A 0R6, Canada}
 \address{Department of Physics, University of Ottawa, Ottawa, Ontario, K1N 6N5, Canada}

\begin{abstract}
We propose and experimentally demonstrate a scheme to generate optically-controlled delays based on off-resonant Raman absorption. Dispersion in a transparency window between two neighboring, optically-activated Raman absorption lines is used to reduce the group velocity of broadband \unit[765]{nm} pulses. We implement this approach in a potassium titanyl phosphate (KTP) waveguide at room temperature, and demonstrate Raman-induced delays of up to \unit[140]{fs} for a 650-fs duration, 1.8-THz bandwidth, signal pulse; the available delay-bandwidth product is $\approx1$. Our approach is applicable to single photon signals, offers wavelength tunability, and is a step toward processing ultrafast photons.
\end{abstract}

\maketitle

The effective operation of communication networks requires the ability to buffer and control the movement of information, whether in a postal service~\cite{Lewins1865}, or a quantum network~\cite{Nature.453.1023}. All-optical controls are particularly desirable for classical and quantum photonic communication systems because manipulations can be effected rapidly, enabling high-bandwidth operations. The need for photonic propagation controls has motivated much research toward the development of slow light devices. These technologies modify the dispersion relationship between photon energy and momentum to reduce the group velocity of optical pulses; for example, using dispersion close to a reflection resonance in periodically-structured photonic media, or material dispersion associated with optical resonance features causing gain or loss~\cite{JOSAB.22.1062}. Potential network applications of slow light include optical switching, signal synchronization, and buffering for all-optical routers~\cite{Nature.433.811}. In addition, slow light offers the prospect of improved optical sensing~\cite{SensorsActuatorsB.173.28}, and enhanced nonlinear interactions~\cite{AdvOptPhoton.2.287,OptLett.21.1936,PhysRevLett.82.5229}. 

Periodically-structured media slow light near reflection resonances, where energy is transferred between strongly-coupled forward- and backward-propagating waves, thereby reducing the group velocity~\cite{AdvOptPhoton.2.287}. Bragg gratings offer a simple implementation, however, more sophisticated coupled resonator structures are necessary to reduce the group velocity while minimizing higher-order dispersion which distorts the shape of pulses. For example, cascaded Bragg gratings~\cite{IEEMicroGWL.8.327}, Moir\'e gratings~\cite{PhysRevA.62.013821,ElecLetters.41.1075}, ring resonators~\cite{IEEEPhotonTechLett.10.994}, and photonic crystal structures~\cite{PhysRevLett.87.253902,PhysRevLett.94.073903,Nature.438.65,OptExpress.16.9245,NatPhoton.2.465} have all been used to demonstrate slow light. However, such structures offer limited tunability and lack continuous optical control when compared with optical absorption and gain resonance techniques.

Slow light has been achieved for pulses tuned between absorbing resonances in an atomic vapour~\cite{PhysRevA.73.063812,PhysRevLett.98.153601,JPhysB.41.051004,NatPhoton.5.230,arxiv.1505.04071}. In one demonstration, pulses as short as \unit[275]{ps} were delayed by up to \unit[6.8]{ps}~\cite{PhysRevLett.98.153601}. Motivated by the desire for optical control of slow light, a variety of nonlinear phenomena have been exploited, including, stimulated Brillouin scattering (SBS)~\cite{PhysRevLett.94.153902,ApplPhysLett.87.081113}, stimulated Raman scattering (SRS)~\cite{OptExp.13.6092}, and electromagnetically-induced transparency (EIT)~\cite{PhysRevLett.66.2593,Hau1999}. In EIT, a control field opens a narrow transparency window within an absorption line. The steep dispersion in the transparency window strongly reduces the group velocity of probe radiation, and the resulting slow light can be used to produce efficient nonlinear interactions~\cite{OptLett.21.1936,PhysRevLett.82.5229}; however, linewidth limitations typically restrict the use of EIT to narrowband pulses with MHz bandwidth~\cite{fleischhauer:633}. Alternative methods which use gain features arising from SBS~\cite{PhysRevLett.94.153902} and SRS~\cite{OptExp.13.6092} have been demonstrated with higher-bandwidth pulses, with durations of \unit[15]{ns} and \unit[430]{fs}, respectively. Techniques which use gain can produce noise photons; such methods are therefore unsuitable for quantum information, and communication, protocols which rely on photon counting.

\begin{figure*}
\centering
\scalebox{0.9}{\includegraphics*[viewport=50 190 550 370]{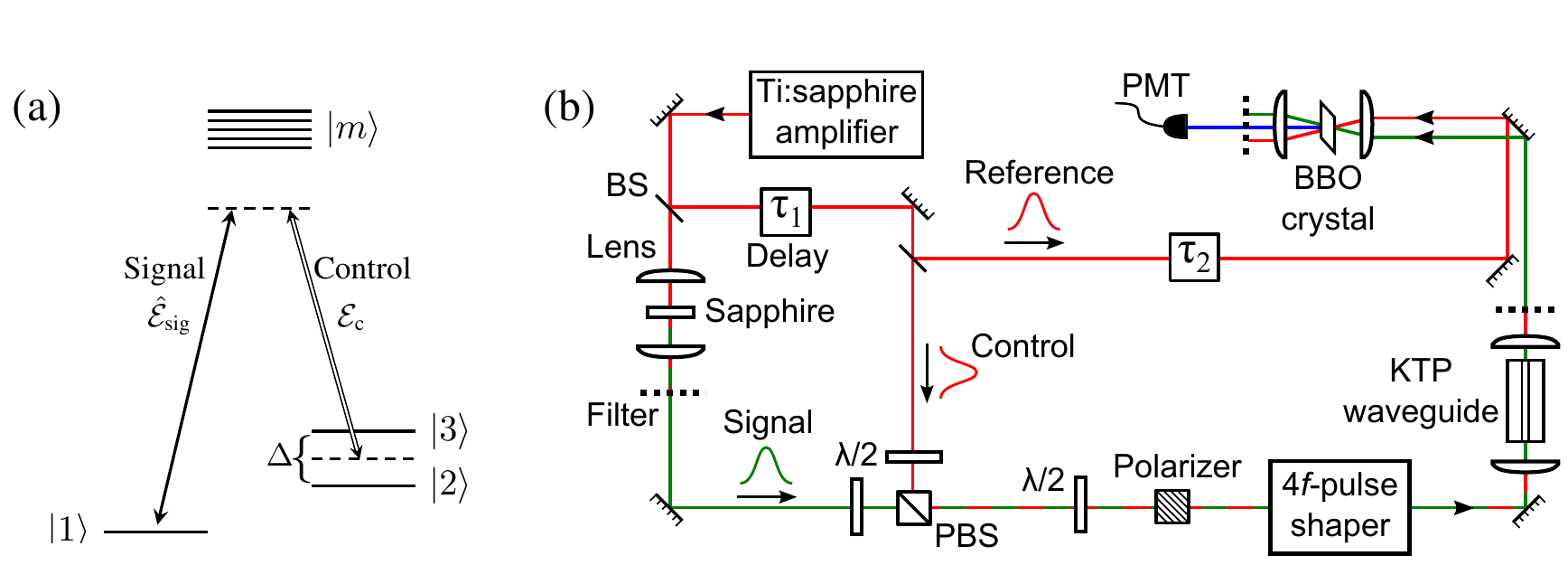}}
\caption{(Color online) Diagram of (a) the generic level configuration, and (b) the experimental layout, of our scheme. The signal field is coupled near two-photon resonance with two Raman-accessible states ($|2\rangle$ and $|3\rangle$).\label{fig:exptlayout}}
\end{figure*}

In this letter, we propose and experimentally demonstrate an all-optical scheme for slow light, suitable for ultra-broadband single photons. Our technique is broadly tunable, and can be implemented in molecular ensembles, atomic vapours, and solids. The scheme reduces the group velocity of ultra-broadband \textsl{signal} photons using off-resonant Raman absorption.  As shown in Fig.~\ref{fig:exptlayout}(a), the signal pulse of frequency $\omega_{\text{sig}}$ copropagates with a strong, near-constant intensity, \textsl{control} pulse of frequency $\omega_{\text{c}}$ through a medium with a double-$\Lambda$-level structure. 
Dipole transitions between the three lower states and the upper manifold of states $|m\rangle$ are permitted, resulting in Raman coupling of levels $\ket{1}\rightarrow\ket{2}$ and $\ket{1}\rightarrow\ket{3}$ when the control and signal pulses are applied. The gradient in linear dispersion between two absorption lines can be used to reduce or increase the group velocity of light \cite{PhysRevA.73.063812, JPhysB.41.051004}. Here, this is achieved for the signal field by setting $(\omega_{\text{sig}}-\omega_{\text{c}})=\omega_2-\omega_1+\Delta/2=\omega_3-\omega_1-\Delta/2$,   where $\hbar\Delta$ is the energy splitting of levels $\ket{2}$ and $\ket{3}$, and $\hbar\omega_{k}$ is the energy of level $\ket{k}$. The control and signal pulses have bandwidths $\delta\omega_{\text{c}}$ and $\delta\omega_{\text{sig}}$, respectively, satisfying $\delta\omega_{\text{c}}\ll\delta\omega_{\text{sig}}<\Delta$  such that the control pulse intensity is near-constant for the duration of the signal pulse. The signal pulse experiences group velocity dispersion from each of the two Raman absorption lines, but absorption losses are minimized by the detuning from two-photon resonance. This Raman scheme enables fast control and broad tunability of the dispersion by switching the intensity and wavelength of the control pulse. In contrast to slow light techniques which involve gain in an amplifier, we expect this scheme to be suitable for controlling single photon pulses while maintaining their quantum character. The speed and quantum-compatibility of this method make it an attractive tool for pulse sequencing in quantum key distribution, for example~\cite{Nature.509.475,PhysRevLett.114.180502}.

The layout of our experiment is shown in Fig.~\ref{fig:exptlayout}(b). The signal and control pulses are generated using a Coherent RegA Ti:sapphire laser system which outputs {160-fs} pulses, centered at \unit[795]{nm}, at a repetition rate of \unit[49]{kHz}. The beam is partitioned at a beam splitter (BS) and the transmitted beam is focused into a sapphire plate to generate a white light continuum spectrum by self-phase modulation~\cite{ApplPhysB.97.561}; the white light spectrum is spatially- and spectrally-filtered to create the signal pulse for the slow light interaction. The beam reflected by the beam splitter is sent to an optical delay line ($\tau_1$). This beam is subsequently partitioned to create a control pulse for the slow light interaction and a \textsl{reference} pulse; the reference pulse, duration $\tau_{\text{ref}}\approx\unit[160]{fs}$, is sent to a cross-correlator with variable delay ($\tau_2$) for measurement of the slow light. The signal and control beams are combined at a polarizing beam splitter (PBS) and sent through an amplitude 4$f$ pulse shaper which allows independent spectral control of each beam by translating knife edges in the focal plane of the shaper~\cite{weiner2009ultrafast}. We spectrally-filter the control spectrum to $<\unit[1]{nm}$ bandwidth, centered at $\lambda_{\text{c}}=\unit[795]{nm}$; the control pulse has a full-width at half maximum (FWHM) duration of $\tau_{\text{c}}=\unit[4]{ps}$. With no filtering in the 4$f$ shaper, the signal spectrum spans the range $\unit[755]{nm}\lesssim\lambda_{\text{sig}}\lesssim\unit[780]{nm}$. After filtering, the signal pulse has a FWHM bandwidth of \unit[3.6]{nm}, corresponding to a Fourier-limited pulse duration of $\tau_{\text{sig}}^{\text{FL}}\approx\unit[490]{fs}$. The control pulse, energy $\mathcal{E}_{\text{c}}=\unit[14.3(8)]{nJ}$, and signal pulse, energy $\mathcal{E}_{\text{sig}}=\unit[40(1)]{pJ}$,  are then focussed into a \unit[3]{cm}-long potassium titanyl phosphate (KTP) single-mode waveguide~\cite{advr,AppPhysLett.104.051117}. At the output of the waveguide, the beam is collimated and sent to a spectrometer (Ocean Optics HR4000), or to a background-free, non-collinear, cross-correlation setup~\cite{weiner2009ultrafast}. The cross-correlation setup allows measurement of the signal pulse arrival time and temporal intensity profile by type-I sum-frequency generation (SFG) with the reference pulse in a 200-$\mu$m-thick beta-barium borate (BBO) crystal. The SFG cross-correlation signal is detected using a photo-multiplier tube (PMT), and boxcar-integrated for data acquisition.

\begin{figure}
\centering
\includegraphics[scale=1]{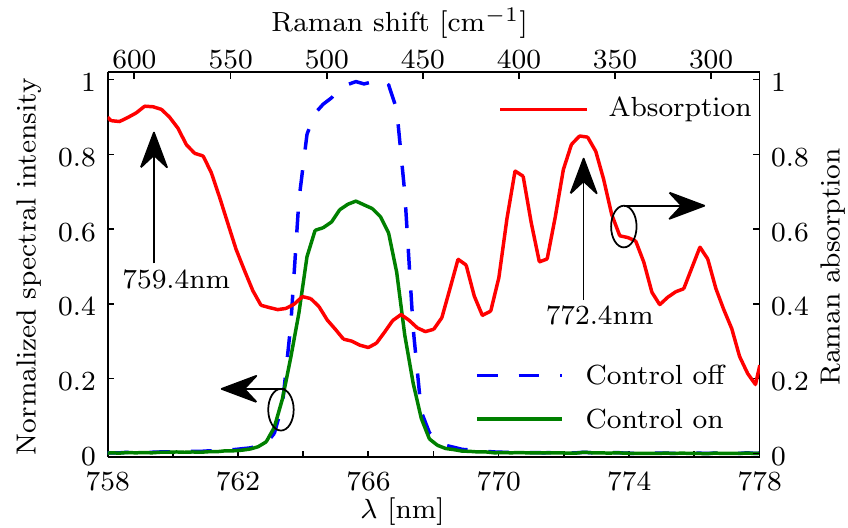} 
\caption{Right-hand ordinate: plot of the measured Raman absorption spectrum of the KTP waveguide (red curve) showing absorption peaks at \unit[759.4]{nm} and \unit[772.4]{nm}. Left-hand ordinate: the signal spectrum, pre-filtered in the 4$f$-shaper, and measured at the output of the waveguide, is plotted with the \unit[795]{nm} control pulse off (blue, dashed curve), and on (green, solid curve), showing that the control pulse causes Raman absorption of the signal pulse.  \label{fig:spectrum}}
\end{figure}

The KTP crystal is Z-cut, with a waveguide written in the X-direction; the mode field-diameter is ${\sim\unit[4]{\mu m}}$. The control and signal pulses propagate in the X-direction, each with Z-polarization. In this configuration [X(ZZ)X], KTP exhibits multiple vibrational Raman peaks in the Stokes spectrum on the interval from $\sim$\unit[50]{cm$^{-1}$} to $\sim$\unit[850]{cm$^{-1}$} shift~\cite{ApplOpt.19.4136,ApplPhysLett.82.325}. In Fig.~\ref{fig:spectrum}, we plot the signal Raman absorption spectrum between \unit[758]{nm} and \unit[778]{nm}, measured by temporally-overlapping the unfiltered signal pulse with the control pulse in the waveguide. The Raman absorption $\mathcal{A}(\lambda)$ is given by $\mathcal{A}(\lambda)=1-I^{\text{on}}_{\text{sig}}(\lambda)/I^{\text{off}}_{\text{sig}}(\lambda)$, where $I^{\text{off~(on)}}_{\text{sig}}(\lambda)$ is the signal spectral intensity at wavelength $\lambda$ output from the waveguide with the control pulse off~(on).
The spectrum shows absorption peaks at \unit[759.4]{nm} and \unit[772.4]{nm}, separated by a `window' of lower absorption in the range \unit[763]{nm}$\lesssim\lambda_{\text{sig}}\lesssim$\unit[768]{nm}. This absorption structure replicates the doublet required for Raman slow light. The $4f$ pulse shaper is used to filter the signal spectrum to fit in the absorption window as shown in Fig.~\ref{fig:spectrum} (control pulse off, dashed, blue curve); when the control pulse is turned on (solid, green curve), the signal intensity is reduced by ${\approx35\%}$ due to Raman absorption.

\begin{figure}[h]
\centering
\includegraphics[scale=1]{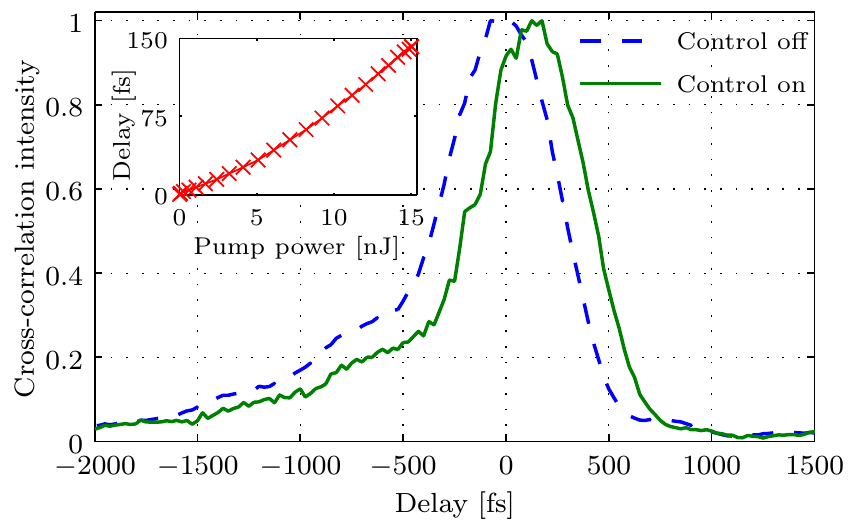} 
\caption{Plot of the normalized cross-correlation intensity between the reference pulse and the signal pulse output from the waveguide with the control pulse off (dashed, blue curve) and on (solid, green curve). The signal pulse is delayed relative to the reference pulse due to Raman interaction with the control pulse. Inset: a plot of the signal delay as a function of control power, measured using the first moment of the cross-correlation intensity $\langle\tau\rangle$; we measure delays of up to \unit[140]{fs}.\label{fig:xcorrelation}}
\end{figure}

In Fig.~\ref{fig:xcorrelation}, we plot the normalized intensity cross-correlation of the signal pulse with the control pulse on and off. The cross-correlation has a FWHM duration of $\tau_{\text{xc}}=\unit[675]{fs}$, giving a signal pulse duration of ${\tau_{\text{sig}}\approx\left(\tau_{\text{xc}}^2-\tau_{\text{ref}}^2\right)^{1/2}=\unit[650]{fs}}$, such that $\tau_{\text{sig}}>\tau_{\text{sig}}^{\text{FL}}$ because the signal is chirped by dispersion in the waveguide. Figure~\ref{fig:xcorrelation} shows that the normalized cross-correlation with the control pulse on (green, solid curve) is delayed in time relative to the case with the control pulse off (blue, dashed curve). The mean pulse delay can be measured by the first moment, or mean, of the cross-correlation intensity,
\[
\langle\tau\rangle=\left.\frac{\int\tau I(\tau)\text{d}\tau}{\int I(\tau)\text{d}\tau}\right\rvert_{\text{control on}}-\left.\frac{\int\tau I(\tau)\text{d}\tau}{\int I(\tau)\text{d}\tau}\right\rvert_{\text{control off}},
\]
where $I(\tau)$ is the cross-correlation intensity as a function of the delay $\tau$ between the signal and reference pulses. In the inset of Fig.~\ref{fig:xcorrelation}, we plot the mean pulse delay $\langle\tau\rangle$ as a function of the input control pulse energy, showing a smooth increase in the delay from zero up to \unit[140]{fs} at the maximum control pulse energy of \unit[14.3(8)]{nJ}.

We model our scheme assuming an ensemble system with the level configuration shown in Fig.~\ref{fig:exptlayout}(a) with two Raman-accessible states ($|2\rangle$ and $|3\rangle$). The signal is optically-coupled to these states, and the two-photon transition is tuned to a point between the two Raman absorption lines. By adiabatic elimination of the high-energy $|m\rangle$-states, and assuming that dipole transitions $\ket{1}\leftrightarrow\ket{2}$ and $\ket{1}\leftrightarrow\ket{3}$ are negligible, we derive the following Maxwell-Bloch equations for this system~\cite{PhysRevA.24.1980, PhysRevA.73.023810,supplementary},
\begin{align}\label{MaxwellBloch1}
&{\dot{\hat Q}}_{31}(z,t)=-\Gamma_3{\hat Q}_{31}(z,t)-i\kappa_{13} E_{\text{c}}^*(z,t){\hat E}_{\text{sig}}(z,t)e^{-i\frac{\Delta}{2}t}\\
&{\dot {\hat Q}}_{21}(z,t)=-\Gamma_2{\hat Q}_{21}(z,t)-i\kappa_{12} E_{\text{c}}^*(z,t){\hat E}_{\text{sig}}(z,t)e^{i\frac{\Delta}{2}t}, 
\end{align}
\begin{align}\label{MaxwellBloch2}\nonumber
\left(\partial_z+\frac{1}{c}\partial_t \right){\hat E}_{\text{sig}}(z,t)&=i E_{\text{c}}(z,t) [ \beta_{31} {\hat Q}_{31}(z,t)e^{i\frac{\Delta}{2}t} \\ 
&+ \beta_{21} {\hat Q}_{21}(z,t)e^{-i\frac{\Delta}{2}t}].
\end{align}
Here, ${\hat E}_{\text{sig}}(z,t)$, $ E_{\text{c}}(z,t)$ and ${\hat Q}_{ij}(z,t)$ are slowly-varying components, respectively, of the signal field, control field, and coherences in the medium between states $\ket{i}$ and $\ket{j}$, as a function of propagation coordinate $z$ and time $t$; we ignore transverse field effects and consider propagation over a distance $L$. The frequency spacing between the two Raman lines is denoted by $\Delta$, and $\Gamma_{2,3}$ are absorption linewidths; the coupling coefficients are
$\kappa_{ij}=\sum_{m}{d_{im}d_{mj}[(\omega_{mi}-\omega_{\text{sig}})^{-1}+(\omega_{mi}+\omega_{\text{c}}})^{-1}]$ and $\beta_{ij}=2\pi N\hbar \kappa_{ij}/c$, where $N$ is the number density and $d_{ij}=\langle i|\hat{d}|j\rangle/\hbar$ is the matrix element of the dipole operator $\hat{d}$. Note that we have assumed fixed population in the ground and Raman states because the population transfer is negligible with a weak signal field.

Equations (\ref{MaxwellBloch1}-\ref{MaxwellBloch2}) can be solved numerically to demonstrate the Raman-induced group delay. However, the most important features of our scheme can be captured through an analytical solution by assuming a control field with constant intensity. As shown in the supplemental material~\cite{supplementary}, we can analytically find the linear response function and consequently describe the loss and  group delay as a function of the peak optical depth of the medium. Assuming that the absorption lines have the same Raman couplings and linewidths, such that $\kappa_{12}\beta_{21}=\kappa_{13}\beta_{31}$ and $\Gamma=\Gamma_2=\Gamma_3$, respectively, we show that the group delay $\tau_{\text{g}}$ is given by,
\[
\tau_{\text{g}}=d_0\frac{\Gamma(\Delta^2/4-\Gamma^2)}{(\Delta^2/4+\Gamma^2)^2},
\]
with an associated loss in dB of,
\[
\quad\eta=d_0\frac{10}{\text{ln}(10)}\frac{2\Gamma^2}{\Delta^2/4+\Gamma^2},
\]
where $d_0=k_0L\kappa_{12}\beta_{21}|E_{\text{c}}|^2/\Gamma$ is the Raman peak optical depth in resonance with each Raman line, and $k_0$ is the wavevector of the signal centre frequency.

 \begin{figure}[ht]
\scalebox{0.52}{\includegraphics*[viewport=25 218 600 540]{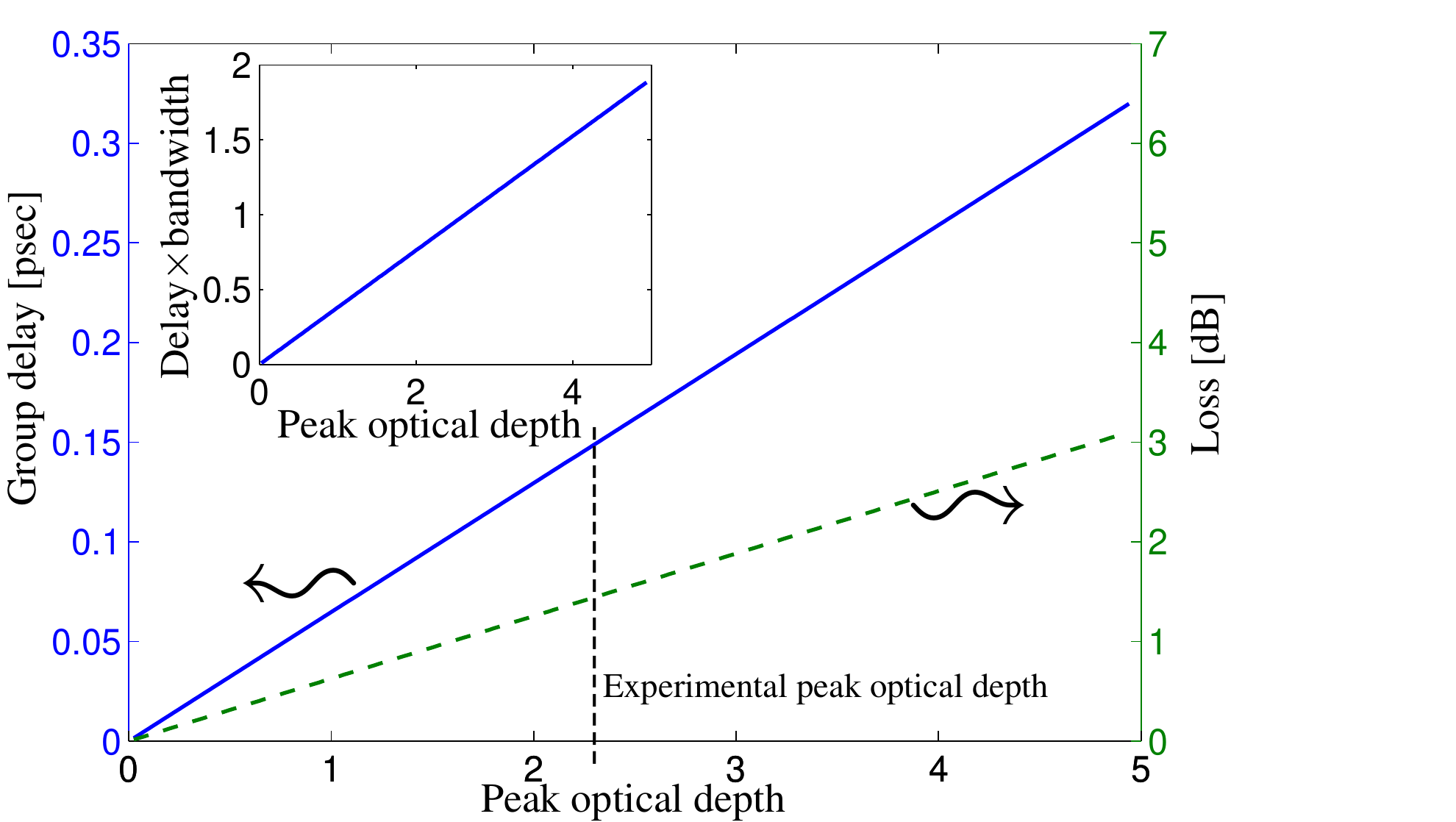}}
\caption{(Color online) Group delay and loss with respect to the Raman peak optical depth. We assume $\Gamma=1$~THz and $\Delta=6.8$~THz, such that the absorption and dispersion based on the model resemble that of the experiment; see Fig.~SI~1 in the supplemental material~\cite{supplementary}. The inset shows the delay-bandwidth product, where the bandwith is $\Delta-\Gamma$. }\label{fig:delay_loss}
\end{figure}

The expected group delay and loss are shown in Fig.~\ref{fig:delay_loss} for $\Gamma=1$~THz and $\Delta=6.8$~THz as a function of the peak optical depth. The parameters $\Delta$ and $\Gamma$ are chosen such that the response function (derived from our theoretical description) matches the measurement results; see Fig.~SI 1 in the supplemental material~\cite{supplementary}. The delay-bandwidth product can be defined as a more explicit measure of our scheme\rq{}s performance. Assuming that the available bandwidth of the transparency window is given by $\Delta-\Gamma$, we plot the theoretical delay-bandwidth product in the inset of Fig.~\ref{fig:delay_loss}. A delay-bandwidth product of 1 means that the delay will be equal to the transform-limited signal duration if all of the available bandwidth is used. For our parameters this can be achieved at $d_0\approx 2.5$, which is demonstrated for one of the Raman absorption peaks in our experiment. 

The measurement results in the inset of Fig.~\ref{fig:xcorrelation} for delay vs. power slightly deviates from our predicted linear dependence shown in Fig.~\ref{fig:delay_loss}. This is because the Lorentzian lineshapes used in the theoretical model result in negligible group delay dispersion (GDD) in the transparency window; however, in the experiment, the more complicated wavelength dependence of the absorption spectrum (see Fig.~\ref{fig:spectrum}) causes GDD which results in the nonlinear dependence of Fig.~\ref{fig:xcorrelation} (inset)~\cite{supplementary}.

In summary, we have introduced a Raman-based scheme for generating optically-induced slow light, and demonstrated it in a KTP waveguide which provided a transparency window with approximately \unit[5.8]{THz} of bandwidth. Using a signal pulse at \unit[765]{nm} with \unit[1.8]{THz} bandwidth, corresponding to a Fourier-limited pulse duration of \unit[490]{fs}, we achieved power-dependent delays of up to 140~fsec.

The operational bandwidth of our scheme is determined by the medium; the delay depends on the control field intensity, frequency spacing between Raman levels, and Raman linewidths. Given that we use linear dispersion between two absorption lines, temporal delay is accompanied by loss. Delay per loss remains independent of the peak optical depth and only depends on the frequency spacing between the Raman absorption lines, and the Raman linewidths. For the theoretical parameters modelled, a delay-bandwidth product of 1 corresponds to $\approx$1.5~dB loss. The loss can be reduced by using a medium with narrower Raman linewidths. 

In principle, due to the use of absorption features, rather than gain features which add noise photons to the delayed signal pulse, our scheme can be implemented to process single-photons. Furthermore, noise photons created by Raman scattering of control photons from thermal excitations could be spectrally-filtered to preserve the single photon charactersitics of the signal. Thanks to the off-resonant Raman coupling, the delay can be controlled optically, delivering frequency tunability, and the possibility of operating on multiple signals concurrently by frequency-multiplexing. Furthermore, alternative media with more Raman lines can be considered to provide multimode functionality with one control field. For example, ro-vibrational manifolds of liquid or gas phase molecular ensembles in hollow-core fibres are attractive candidate systems~\cite{NatPhoton.8.278}. We expect that this scheme will find use in ultrafast quantum processing applications, perhaps in combination with a frequency-shifter~\cite{PhysRevLett.105.093604,FisherPrep2015} for pulse sequencing, for example~\cite{nature.461.241,NJP.16.065019}.

The authors are grateful to Paul Hockett and Rune Lausten for helpful discussions and comments on the manuscript. The authors wish to thank Philip Battle of AdvR Inc.~for selection of the waveguide. This work was partially-supported by NSERC.


\begin{thebibliography}{44}
\bibitem [{adv()}]{advr}%
  \BibitemOpen
  \href@noop {} {}\bibinfo {note} {The waveguide was fabricated by AdvR,
  Inc.}\BibitemShut {Stop}%
\end{thebibliography}
%

\renewcommand{\theequation}{SI\arabic{equation}}

\setcounter{table}{0}
\setcounter{equation}{0}
\setcounter{figure}{0}
\makeatletter
\renewcommand{\bibnumfmt}[1]{[SI#1]}
\renewcommand{\citenumfont}[1]{SI#1}
\renewcommand{\figurename}[1]{FIG. SI#1}

\widetext
\begin{center}
\textbf{\large Supplemental Material for ``Ultrafast slow-light: Raman-induced delay of THz-bandwidth pulses"}

\author{Authors}
\affiliation{National Research Council of Canada, 100 Sussex Drive, Ottawa, Ontario, K1A 0R6, Canada}
\end{center}

\section{Equations of motion} 

We describe here the equations of motion that are used for deriving the Maxwell-Bloch equations~\cite{PhysRevA.24.1980, PhysRevA.73.023810}. The total Hamiltonian ($H_{\text{tot}}$) driving dynamics of the system includes the internal energy of states of the medium and interaction with electromagnetic fields based on the dipole approximation. Using the Heisenberg relation ($\frac{d{\mathcal {\hat O}}}{dt}=\frac{i}{\hbar}\left[H_{\text{tot}}, {\mathcal {\hat O}}\right]+\frac{\partial {\mathcal {\hat O}}}{\partial t}$), one can derive the following set of equations,
\begin{align}
\frac{d{\hat \rho}_{31}}{dt}&=i\omega_{31}{\hat \rho}_{31}+i\sum_{m}{d_{1m}{\mathcal E}_{\text{tot}}{\hat \rho}_{3m}}-i\sum_{m}{d_{m3}{\mathcal E}_{\text{tot}}{\hat \rho}_{m1}}\\ 
\frac{d{\hat \rho}_{21}}{dt}&=i\omega_{21}{\hat \rho}_{21}+i\sum_{m} d_{1m}{\mathcal E}_{\text{tot}}{\hat \rho}_{2m}-i\sum_{m} d_{m2}{\mathcal E}_{\text{tot}}{\hat \rho}_{m1}\\ 
\label{eq:rhom1}\frac{d{\hat \rho}_{m1}}{dt}&=i\omega_{m1}{\hat \rho}_{m1}-id_{1m}{\mathcal E}_{\text{tot}}({\hat \rho}_{11}-{\hat \rho}_{mm})-id_{3m}{\mathcal E}_{\text{tot}}{\hat \rho}_{31} -id_{2m}{\mathcal E}_{\text{tot}}{\hat \rho}_{21}\\ 
\label{eq:rho3m}\frac{d{\hat \rho}_{3m}}{dt}&=i\omega_{3m}{\hat \rho}_{3m}-id_{m3}{\mathcal E}_{\text{tot}}({\hat \rho}_{mm}-{\hat \rho}_{33})+id_{m1}{\mathcal E}_{\text{tot}}{\hat \rho}_{31}\\ 
\label{eq:rho2m}\frac{d{\hat \rho}_{2m}}{dt}&=i\omega_{2m}{\hat \rho}_{2m}-id_{m2}{\mathcal E}_{\text{tot}}({\hat \rho}_{mm}-{\hat \rho}_{22})+id_{m1}{\mathcal E}_{\text{tot}}{\hat \rho}_{21},
\end{align}
where ${\hat \rho}_{ij}=|i\rangle\langle j|$ and ${\mathcal E}_{\text{tot}}={\mathcal E}_{\text{c}}(z,t)+{\hat {\mathcal E}}_{\text{sig}}(z,t)$. The function ${\mathcal E}_{\text{c}}(z,t)=E_{\text{c}}(z,t)e^{i(\omega_{\text{c}} t-k_{\text{c}} z)}+E_{\text{c}}^*(z,t)e^{-i(\omega_{\text{c}} t-k_{\text{c}} z)}$ denotes the control field and ${\hat{\mathcal E}}_{\text{sig}}(z,t)={\hat E}_{\text{sig}}(z,t)e^{i(\omega_{\text{sig}}t-k_{\text{sig}}z)}+{\hat E}^{\dagger}_{\text{sig}}(z,t)e^{-i(\omega_{\text{sig}}t-k_{\text{sig}}z)}$ represents the signal field. The dipole matrix elements are $d_{ij}=\langle i|{\hat d}|j\rangle/\hbar$, and $\omega_{ij}$ is the angular frequency of transition from $|j\rangle$ to $|i\rangle$. As shown in Fig.~1(a), the $|m\rangle$-states represent the excited state manifold.

In order to eliminate the excited state dynamics, we integrate equations (\ref{eq:rhom1}-\ref{eq:rho2m}) from the above set and use the result to rewrite the dynamics of ${\hat \rho}_{31}$ and ${\hat \rho}_{21}$. We neglect the Stark shift term and eliminate the fast-rotating terms compared to slowly-varying components. For a scatterer at position $z$, this results in,
\begin{align}\label{BlochEq1}
\frac{d{\hat \rho}_{31}}{dt}&=i\omega_{31}{\hat \rho}_{31}-i\kappa_{13} E_{\text{c}}^*(z,t){\hat E}_{\text{sig}}(z,t)e^{-i\frac{\Delta}{2}t}\times ({\hat \rho}_{11}-{\hat \rho}_{33})e^{i\omega_{31}t} e^{i(k_{\text{c}}-k_{\text{sig}})z}, \text{ and,}\\  \label{BlochEq2}
\frac{d{\hat \rho}_{21}}{dt}&=i\omega_{21}{\hat \rho}_{21}-i\kappa_{12} E_{\text{c}}^*(z,t){\hat E}_{\text{sig}}(z,t)e^{i\frac{\Delta}{2}t}\times ({\hat \rho}_{11}-{\hat \rho}_{22})e^{i\omega_{21}t}e^{i(k_{\text{c}}-k_{\text{sig}})z},
\end{align}
where,
\begin{equation}\nonumber
\kappa_{1i}=\sum_{m}{d_{1m}d_{mi}\left( \frac{1}{\omega_{m1}-\omega_{\text{sig}}}+\frac{1}{\omega_{m1}+\omega_{\text{c}}}\right)}.
\end{equation}

We rewrite the above equations for slowly-varying components (${\hat Q}_{ij}$) of the matter coherences (${\hat \rho}_{ij}$); this leads to,
\begin{align}\label{BlochEq3}
\frac{d{\hat Q}_{31}}{dt}&=-\Gamma_3{\hat Q}_{31}-i\kappa_{13} E_{\text{c}}^*(z,t){\hat E}_{\text{sig}}(z,t)e^{-i\frac{\Delta}{2}t}\times ({\hat Q}_{11}-{\hat Q}_{33}), \text{ and,}\\
\frac{d{\hat Q}_{21}}{dt}&=-\Gamma_2{\hat Q}_{21}-i\kappa_{12} E_{\text{c}}^*(z,t){\hat E}_{\text{sig}}(z,t)e^{i\frac{\Delta}{2}t}\times ({\hat Q}_{11}-{\hat Q}_{22}),
\end{align}
where ${\hat Q}_{\nu1}(t)= {\hat \rho}_{\nu1}e^{-i\omega_{\nu1}t}e^{-i(k_{\text{c}}-k_{\text{sig}})z}$ for $\nu=2,3$ and $\Gamma_{\nu}$ are linewidths associated to $|\nu\rangle$ Raman lines. We assume a cylindrically-shaped ensemble of length $L$ along the $z$ (propagation) direction and we define collective operators by averaging over all scatterers with a slice of length $\Delta z$ with $N_z$ atoms (or molecules). Assuming that $N_z\gg 1$ and that all of the scatterers are initially in state $|1\rangle$, we can simplify the above Bloch equations to
\begin{align}\label{BlochEq3}
&{\dot{\hat Q}}_{31}(z,t)=-\Gamma_3{\hat Q}_{31}(z,t)-i\kappa_{13} E_{\text{c}}^*(z,t){\hat E}_{\text{sig}}(z,t)e^{-i\frac{\Delta}{2}t}\\
&{\dot {\hat Q}}_{21}(z,t)=-\Gamma_2{\hat Q}_{21}(z,t)-i\kappa_{12} E_{\text{c}}^*(z,t){\hat E}_{\text{sig}}(z,t)e^{i\frac{\Delta}{2}t}. 
\end{align}
Here, ${\hat Q}_{ij}(z,t)=\sum_{k=1}^{N_z}{{\hat Q}_{ij}^{(k)}(t)}$ and ${\hat Q}_{ij}^{(k)}(t)$ is the coherence operator for the $k^{th}$ atom in a slice of the ensemble at position $z$.

\subsection{ Propagation of EM fields} Propagation of the signal and control fields are determined by the total macroscopic polarization that these fields experience in the medium. We are interested in single-photon (weak) signal fields. The signal field propagation is therefore given by the 1-dimensional wave equation
\begin{equation}\label{Maxwell1}
\left(\partial^2_z-\frac{1}{c^2}\partial^2_t \right){\hat {\mathcal E}_{\text{sig}}}(z,t)=\frac{4\pi}{c^2}\partial^2_t{\hat {\mathcal P}_{\text{sig}}(z,t)}, 
\end{equation}
where 
\begin{equation}
{\hat {\mathcal P}_{\text{sig}}(z,t)}=\frac{\hbar}{A\Delta z}\sum_{i=1}^{N_z}\sum_{m}{\left[ d_{1m} {\hat \rho}_{1m}^i + d_{m3} {\hat \rho}_{m3}^i + d_{m2} {\hat \rho}_{m2}^i\right]}.
\end{equation}
Similar to the derivation for equations (\ref{BlochEq1}) and (\ref{BlochEq2}), and by keeping only terms that oscillate near the frequency of $\omega_{\text{sig}}$, we find,
\begin{eqnarray}
\nonumber
{\hat {\mathcal P}_{\text{sig}}(z,t)}&=N\hbar E_{\text{c}}(z,t) \left[ \kappa_{31} {\hat Q}_{31}(z,t)e^{i\frac{\Delta}{2}t} + \kappa_{21} {\hat Q}_{21}(z,t)e^{-i\frac{\Delta}{2}t}\right]\\ 
&\times e^{i(\omega_{\text{sig}}t-k_{\text{sig}}z)},
\end{eqnarray}
where $N$ is the number density per unit volume. One can simplify Eq.~(\ref{Maxwell1}) to the following first order equation for the forward propagating component of the signal field,
\begin{eqnarray}\label{Maxwell2}\nonumber
\left(\partial_z+\frac{1}{c}\partial_t \right){\hat  E}_{\text{sig}}(z,t)&=i E_{\text{c}}(z,t) \left[ \beta_{31} {\hat Q}_{31}(z,t)e^{i\frac{\Delta}{2}t} + \beta_{21} {\hat Q}_{21}(z,t)e^{-i\frac{\Delta}{2}t}\right], 
\end{eqnarray}
where $\beta_{i1}=\frac{2\pi N\hbar \kappa_{i1}}{c}$ for $i=\{2,3 \}$.

We now convert the operators to single excitation wavefunctions, with ${\hat  E}_{\text{sig}}(z,t)\rightarrow{E}_{\text{sig}}(z,t)$, ${\hat Q}_{21}(z,t)\rightarrow{Q}_{21}(z,t)$, and ${\hat Q}_{31}(z,t)\rightarrow{Q}_{31}(z,t)$. Assuming a control field with fixed intensity, applicable for a control field that is long compared to the signal duration, we can analytically solve Eqs.~(1-3) by applying a Fourier transformation $\mathcal{F}$ to each equation. This results in,
\begin{eqnarray}
\left(\partial_z+i\frac{\omega}{c}\right){\tilde E}_{\text{sig}}(z,\omega)&=&iE_{\text{c}} \beta_{31} {\tilde Q}_{31}(z,\omega-\Delta/2)+iE_{\text{c}} \beta_{21} {\tilde Q}_{21}(z,\omega+\Delta/2),\label{FT_Maxwell1}\\ 
{\tilde Q}_{31}(z,\omega-\Delta/2)&=&\frac{i\kappa_{13}E_{\text{c}}^*}{i(\omega-\Delta/2)+\Gamma_3}{\tilde E}_{\text{sig}}(z,\omega)\text{, and}\label{FT1},\\
{\tilde Q}_{21}(z,\omega+\Delta/2)&=&\frac{i\kappa_{12}E_{\text{c}}^*}{i(\omega+\Delta/2)+\Gamma_2}{\tilde E}_{\text{sig}}(z,\omega)\label{FT2},
\end{eqnarray}
where ${\tilde E}_{\text{sig}}(z,\omega)=\mathcal{F}\{{E}_{\text{sig}}(z,t)\}$, ${\tilde Q}_{31}(z,\omega-\Delta/2)=\mathcal{F}\{Q_{31}(z,t)e^{i\frac{\Delta}{2}t} \}$, and ${\tilde Q}_{21}(z,\omega+\Delta/2)=\mathcal{F}\{Q_{21}(z,t)e^{-i\frac{\Delta}{2}t} \}$. Using equations (\ref{FT1}) and (\ref{FT2}), one can rewrite Eq.~(\ref{FT_Maxwell1}) as 
\begin{equation}\label{FT_Maxwell2}
\left(\partial_z+i\frac{\omega}{c}\right){\tilde E}_{\text{sig}}(z,\omega)=ik_0\chi(\omega){\tilde E}_{\text{sig}}(z,\omega),
\end{equation}
where 
\begin{equation}
\chi(\omega)=\frac{i}{k_0}\left( \frac{c_{13}}{i(\omega-\Delta/2)+\Gamma_3}+\frac{c_{12}}{i(\omega+\Delta/2)+\Gamma_2}\right),
\end{equation}
is the linear susceptibility of the Raman transitions, $k_0$ is the wavevector associated to the central wavelength of the signal, $c_{13}=\kappa_{13}\beta_{31}|E_{\text{c}}|^2$, and $c_{12}=\kappa_{12}\beta_{21}|E_{\text{c}}|^2$. The expected linear dispersion between the two Raman absorption lines can be described by,
\begin{equation}
\chi(\omega)\approx\chi(0)+\omega\frac{d\chi}{d\omega}+....
\end{equation}
Assuming $\Gamma=\Gamma_2=\Gamma_3$ and $c_1=c_{12}=c_{13}$, this results in group delay of,
\begin{equation}
\tau_{\text{g}}=\frac{L}{v_\text{g}}-\frac{L}{c}=\frac{k_0L}{2}\frac{d\chi}{d\omega}=\frac{k_0 L}{2}\frac{2c_{1}(\Delta^2/4-\Gamma^2)}{(\Gamma^2+\Delta^2/4)^2},
\end{equation}
where $v_\text{g}$ is the group velocity; the commensurate loss, in dB, is,
\begin{equation}
\eta=\frac{10 k_0 L}{\text{ln}(10)}\frac{2c_{1}\Gamma}{\Gamma^2+\Delta^2/4}.
\end{equation}
Consequently, this leads to a constant delay per dB loss of,
\begin{equation}
\frac{\tau}{\eta}=\frac{\text{ln}(10)}{20}\frac{\Delta^2/4-\Gamma^2}{\Gamma(\Gamma^2+\Delta^2/4)}.
\end{equation}

\subsection{Absorption and Dispersion}
\begin{figure}[ht]
\scalebox{0.7}{\includegraphics*[viewport=50 200 530 500]{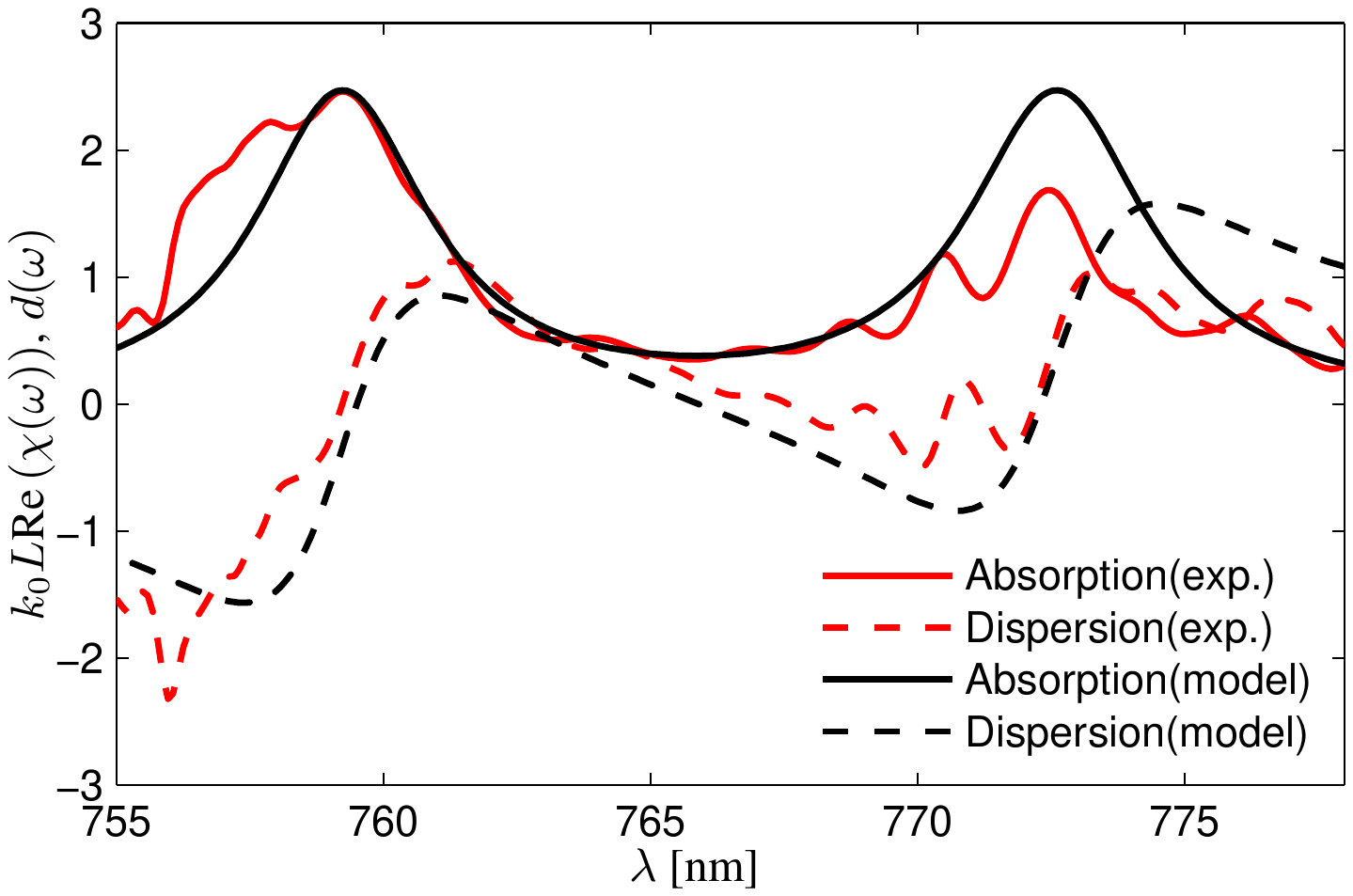}}
\caption{(Color online) Plot of measured Raman-induced optical depth of the KTP waveguide (red curve). The measured $\alpha(\omega)=k\text{Im}\bm{\left(}\chi(\omega)\bm{\right)}$ is used to  numerically calculate $\text{Re}\bm{\left(}\chi(\omega)\bm{\right)}$ via the Kramers-Kronig relation (red dashed curve), where $k$ denotes the wavevector associated to the central wavelength of the signal. The solid (dashed) lines are to show the imaginary (real) part of the response function based on our model; see text for more detail.}\label{chi}
\end{figure}

In addition to our model based on two Raman absorption lines, we can numerically evaluate the real part of the medium's response function. Given that the Raman-induced optical depth can be defined as $d(\omega)=\alpha(\omega)L=-\text{ln}\left( I_{\text{sig}}^{\text{on}}/I_{\text{sig}}^{\text{off}}\right)$, and $\alpha(\omega)L=k_0L\text{Im}\bm{\left(}\chi(\omega)\bm{\right)}$ we can use the Kramers-Kronig relation to evaluate $\text{Re}\bm{\left(}\chi(\omega)\bm{\right)}$; see Fig.~SI~\ref{chi}. 

\begin{figure}[ht]
\scalebox{0.7}{\includegraphics*[viewport=50 200 530 520]{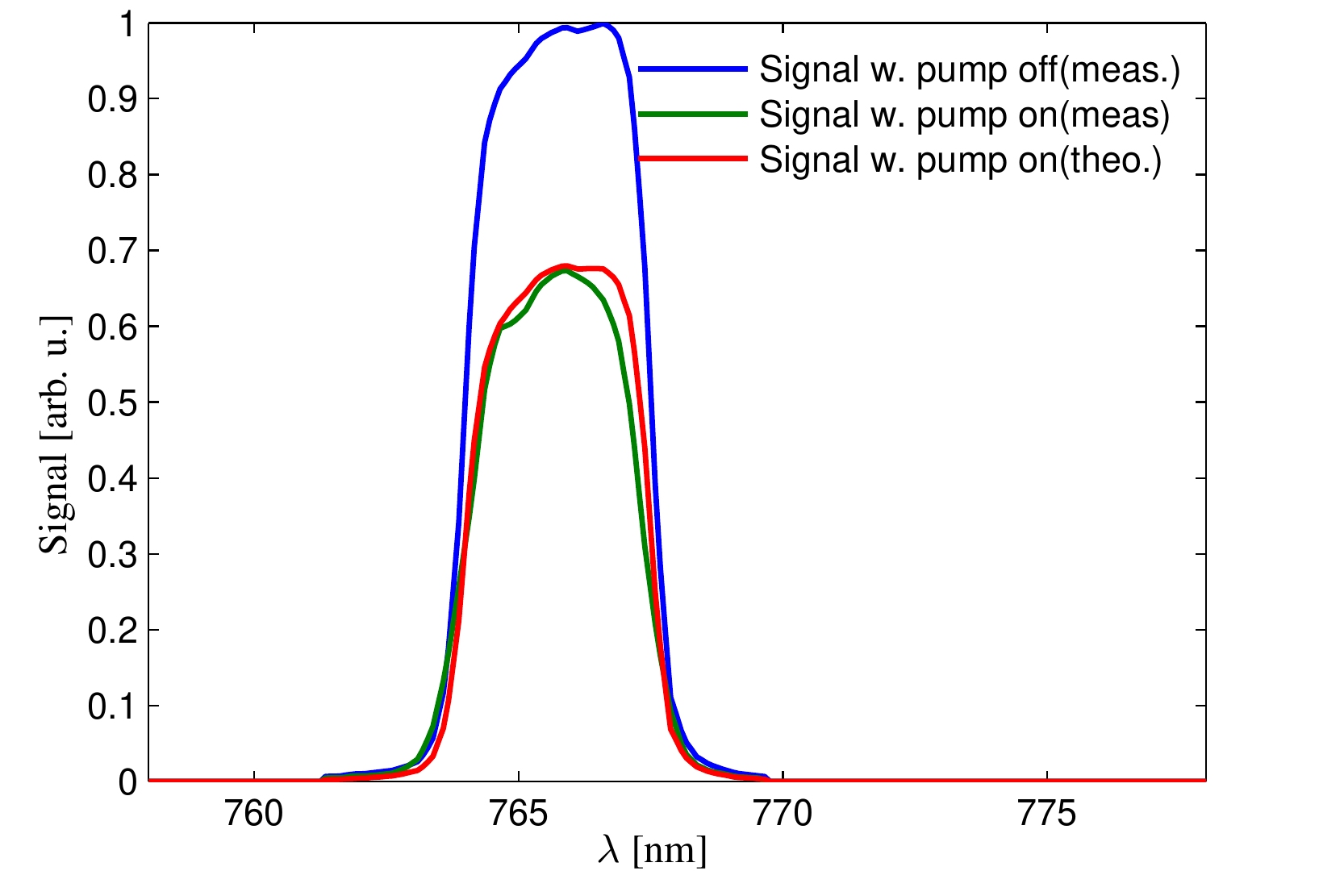}}
\caption{(Color online) Plot the output signal spectrum with the control field off (blue curve) and on (green curve). The red curve is a result of the linear propagation of the signal with the control field off through a medium with a Raman-induced response of $\chi(\omega)$ that has been shown in Fig.~SI~\ref{chi}. }\label{sig_freq}
\end{figure}

Using the Raman-induced susceptibility $\chi(\omega)$, which is shown in Fig.~SI~\ref{chi}, we can evaluate the signal field in the frequency domain. The frequency component of the signal field after propagation is given by ${\tilde E}_{\text{sig}}^{\text{on}}(\omega)={\tilde E}_{\text{sig}}^{\text{off}}(\omega) e^{iL\left[\omega/c+k_0\chi(\omega)/2\right]}$ where ${\tilde E}_{\text{sig}}^{\text{off~(on)}}$ is the signal spectral field strength with the control pulse off~(on). This is shown in Fig.~SI~\ref{sig_freq}, which is in good agreement with the experimental results.

\end{document}